\documentclass[twocolumn, switch]{article} 
\usepackage{preprint}
\usepackage{graphicx}
\usepackage{subcaption}
\usepackage{amsmath}  
\usepackage{booktabs} 
\usepackage{color}
\usepackage{subcaption}
\usepackage{amssymb}  
\usepackage[authoryear]{natbib}

\usepackage[utf8]{inputenc}	
\usepackage[T1]{fontenc}	
\usepackage{xcolor}		
\usepackage[colorlinks = true,
            linkcolor = purple,
            urlcolor  = blue,
            citecolor = cyan,
            anchorcolor = black]{hyperref}	
\usepackage{booktabs} 		
\usepackage{nicefrac}		
\usepackage{microtype}		
\usepackage{lineno}		
\usepackage{float}			

\usepackage{lipsum}		

\usepackage{newfloat}
\DeclareFloatingEnvironment[name={Supplementary Figure}]{suppfigure}
\usepackage{sidecap}
\sidecaptionvpos{figure}{c}

\usepackage{titlesec}
\titlespacing\section{0pt}{12pt plus 3pt minus 3pt}{1pt plus 1pt minus 1pt}
\titlespacing\subsection{0pt}{10pt plus 3pt minus 3pt}{1pt plus 1pt minus 1pt}
\titlespacing\subsubsection{0pt}{8pt plus 3pt minus 3pt}{1pt plus 1pt minus 1pt}

\usepackage{tikz,xcolor,hyperref}

\definecolor{lime}{HTML}{A6CE39}
\DeclareRobustCommand{\orcidicon}{
	\begin{tikzpicture}
	\draw[lime, fill=lime] (0,0) 
	circle [radius=0.16] 
	node[white] {{\fontfamily{qag}\selectfont \tiny ID}};
	\draw[white, fill=white] (-0.0625,0.095) 
	circle [radius=0.007];
	\end{tikzpicture}
	\hspace{-2mm}
}
\foreach \x in {A, ..., Z}{\expandafter\xdef\csname orcid\x\endcsname{\noexpand\href{https://orcid.org/\csname orcidauthor\x\endcsname}
			{\noexpand\orcidicon}}
}

\title{Varying Horizon Learning Economic MPC With Unknown Costs of Disturbed Nonlinear Systems}

\usepackage{xwatermark}
\newwatermark[firstpage,color=gray!90,angle=0,scale=0.28, xpos=0in,ypos=-5in]{*correspondence: \texttt{hdfzj@zjut.edu.cn}}

\usepackage{authblk}

\author[1\thanks{\tt{1112103020@zjut.edu.cn}}]{Weiliang Xiong\orcidA{}}
\author[1\thanks{\tt{hdfzj@zjut.edu.cn}}]{Defeng He}
\author[2\thanks{\tt{hdu@uow.edu.au}}]{Haiping Du}
\author[1\thanks{\tt{jianbinmu@zjut.edu.cn}}]{Jianbin Mu}

\affil[1]{College of Information Engineering, Zhejiang University of Technology, Hangzhou 310023, PR China}
\affil[2]{School of Electrical, Computer and Telecommunications Engineering, University of Wollongong, Wollongong NSW 2522, Australia}

\begin{document}

\twocolumn[ 
  \begin{@twocolumnfalse} 
  
\maketitle

\begin{abstract}
This paper proposes a novel varying horizon economic model predictive control (EMPC) scheme without terminal constraints for constrained nonlinear systems with additive disturbances and unknown economic costs. The general regression learning framework with mixed kernels is first used to reconstruct the unknown cost. Then an online iterative procedure is developed to adjust the horizon adaptively. Again, an elegant horizon-dependent contraction constraint is designed to ensure the convergence of the closed-loop system to a neighborhood of the desired steady state. Moreover, sufficient conditions ensuring recursive feasibility and input-to-state stability are established for the system in closed-loop with the EMPC. The merits of the proposed scheme are verified by the simulations of a continuous stirred tank reactor and a four-tank system in terms of robustness, economic performance and online computational burden.
\end{abstract}
\vspace{1.5em}
  \end{@twocolumnfalse} 
] 



\section{INTRODUCTION}
Recently, economic model predictive control (EMPC) has received much attention in academic and industrial communities due to its ability to explicitly cope with constraints, economic optimization and feedback control in the optimal control framework \citep{ellis_tutorial_2014, rawlings_model_2017}. Similar to tracking MPC, at each time EMPC solves a finite-horizon economic optimal control (FHEOC) problem to derive the economically optimal control actions, the first one of which is then implemented into plants in the manner of receding horizon. By means of dissipativity conditions, control Lyapunov functions or multi-objective control, several EMPC schemes with guaranteed stability have been proposed for nominal systems by, e.g., \citet{angeli2012average, alamo2014gradient, muller_economic_2016,ellis_tutorial_2014,he_stability_2015,he2016economic,zavala2015multiobjective}
and references therein.

In practical control scenarios, however, various disturbances may deteriorate the control performance, feasibility and stability of EMPC for nominal systems. Although nominal EMPC has inherent robustness to small disturbances \citep{muller2014necessity}, it is necessary to find EMPC with guaranteed properties at the expense of some conservativeness. To address this issue, several robust EMPC strategies have been developed for uncertain systems. One popular development has been tube-based robust EMPC for linear systems with bounded disturbances due to its manageable computational complexity \citep{bayer2014tube,lejarza2021economic,dong2020homothetic}. In \citet{bayer2016robust}, stochastic information was used to improve the robust performance of the linear EMPC algorithm. To establish provable properties, \citet{dong2020homothetic} and \citet{lejarza2021economic} developed robust EMPC for linear disturbed systems satisfying dissipativity assumptions. In  \citet{villanueva2020set}, a set-dissipativity-based tube EMPC approach was designed for disturbed nonlinear systems. For nonlinear systems with infrequent disturbances, \citet{mcallister2023suboptimal} proposed a robust EMPC with suboptimal performance under the assumption of feasibility. Moreover, using the multi-objective control framework \citep{he_stability_2015}, in \citet{defeng2019input} and \citet{xiong2025learning}, we proposed tube-based robust EMPC with guaranteed feasibility and input-to-state stability (ISS) of disturbed nonlinear systems. In this framework, the conventional dissipativity assumptions can be circumvented in robust EMPC. 

In EMPC, the economic cost is one of the key components for its control performance, feasibility and stability. Hence, it is often explicitly characterized prior to designing the EMPC of a system. However, due to complex market fluctuations and environmental uncertainties, the economic cost is generally time-varying or even unknown in practice. For time-varying, known economic costs, \citet{ferramosca_economic_2014} combined the consistent dissipativity assumption and generalized terminal constraint to design linear EMPC with guaranteed nominal stability. Using equilibrium manifold, \citet{he2021lyapunov}, \citet{ellis2015real} proposed Lyapunov-based stable EMPC of constrained nonlinear systems with changing economic costs. In the multi-objective framework, \citet{he2021lexicographic, wu_stable_2024} established the nominal stability of nonlinear EMPC with changing economic costs.  For unknown economic costs, \citet{kordabad_q-learning_2022} employed Q-learning to train a parameterized storage function satisfying dissipativity conditions of nominal systems. To the best of our knowledge, however, there are no results on the robust EMPC of uncertain nonlinear systems with unknown economic costs.

The online computational burden is another key issue of EMPC for constrained systems, especially with nonlinearities and unknown costs. As the economic cost is usually non-convex and/or non-positive definite, e.g., learned parameterized functions \citep{kordabad_q-learning_2022}, solving an optimization problem of EMPC is more computationally demanding than that of conventional tracking MPC. From the perspective of strategy, the terminal-free and varying horizon ideas are particularly valuable for lessening the computational burden of nonlinear MPC (e.g., \citet{limon_stability_2006, manzano_componentwise_2021, lorenzen_stochastic_2019, sun_dynamic_2022, sun_robust_2019,leung_stability_2024} and references therein). There exist a few studies on EMPC without terminal constraints for disturbance-free systems. For instance, using the dissipativity assumptions, \citet{grune_economic_2013, grune_asymptotic_2014, reble_unconstrained_2012} employed a sufficiently long horizon to compensate terminal constraints, which guarantees the stability of EMPC. In \citet{schwenkel_linearly_2024} and \citet{muller_economic_2016}, this framework was extended to discounted linear EMPC and periodic EMPC, respectively. To avoid the dissipativity verification challenge, \citet{alamir_new_2021} proposed an EMPC with practical stability by introducing state increment weighting for nominal systems with given economic costs. It is notably difficult to determine these weights to achieve the stability of EMPC, especially for disturbed systems with unknown costs.

There are also a few studies on varying horizon EMPC of constrained nonlinear systems. For example, in \citet{ellis_finite-time_2014}, the horizon is reduced by one at each time and asymptotic stability is ensured using the Lyapunov method and a terminal equality constraint. A continuous-time EMPC similar to that in \citet{ellis_finite-time_2014} can be found in \citet{stamouli_recursively_2024}. Based on the relaxed dynamic programming principle and exponential controllability assumption, \citet{grune2010analysis} presented an EMPC with a variable control horizon, in which multiple control actions are applied sequentially \citep{muller_economic_2016}. To mitigate the myopic behavior often caused by shrinking horizons, \citet{xiong2024two} developed a terminal-free varying horizon EMPC algorithm that strikes a balance between the computational burden and long-term performance of EMPC, with known costs and the absence of disturbances.

\begin{figure}
\begin{center}
\includegraphics[height=5cm]{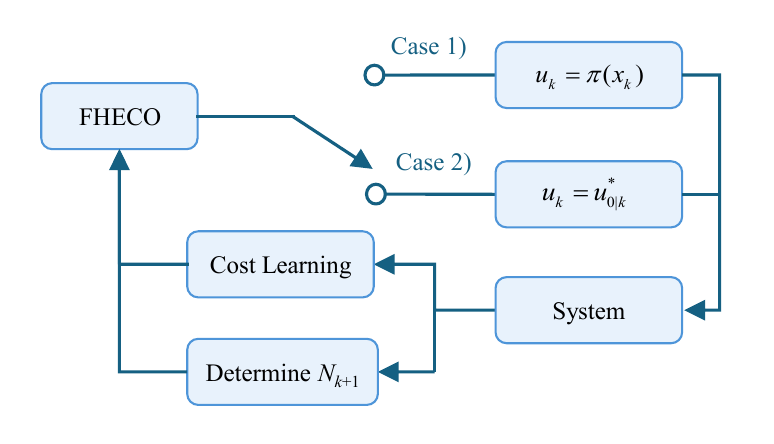}    
\caption{Flowchart of the proposed EMPC scheme} 
\label{fig1}                                 
\end{center}                                 
\end{figure}

In this paper, we develop a novel varying horizon learning robust EMPC scheme without terminal constraints for constrained nonlinear systems subject to bounded disturbances and unknown economic costs.The framework of this scheme is shown in Fig. 1, where at each time $k$, the economic cost is learned using the current measurements and the horizon is determined through a tailored iterative procedure. Then the first action of the optimal control sequence or the terminal control law is applied into the system, depending on the property of the optimal state trajectory. The simulation examples of a continuous stirred tank reactor (CSTR) and a four-tank system validate the performance and advantages of the proposed scheme in terms of control performance, computation burden and robustness.
 
With respect to the available EMPC results, the main contributions of the paper include three parts:  

1) A novel varying horizon learning robust EMPC scheme without terminal constraints is proposed for constrained nonlinear systems subject to bounded disturbances and unknown economic costs. It can be compatible with any regressor for learning unknown economic costs and flexibly makes a trade-off among long-term performance, convergence speed and computational burden via the tailored varying horizon mechanism.
    
2) A general regression learning framework is introduced to online train unknown economic costs and then the combination of the varying horizon mechanism and terminal-free idea is employed to design the nonlinear robust EMPC with unknown costs in the multi-objective control framework.

3) By the tools of robust invariant sets and ISS, some sufficient conditions are established to provably guarantee the recursive feasibility and ISS of the closed-loop EMPC system under various constraints, bounded disturbances and unknown economic costs.

The remainder of this paper is organized as follows. Section II describes the problem and gives preliminaries. Section III presents the new EMPC algorithm, including regressor learning and algorithm design. Section IV analyzes the theoretical properties of the EMPC and two simulation studies are presented in Section V. Section VI concludes the paper. Detailed proofs are provided in the Appendix.
 
\textbf{Notation:} We define $R^{n \times m}$ as $n \times m$-dimensional real space, and $\mathbf{I}_{\geq i}$ is the set of integers not less than $i$, while $\mathbf{I}_a^b = \{a, a+1, \dots, b\}$, $b \geq a$ and $\mathbf{I}_a^b = \emptyset$, $b < a$. Define $|x|$ as the elementwise absolute value and $\|x\|$ as the 2-norm of vector $x$. For two sets $A$ and $B$, the Cartesian product is $A \times B$, the Minkowski addition and Pontryagin difference are $A \oplus B = \{a + b : a \in A,\, b \in B\}$, $A \ominus B = \{a : \{a\} \oplus B \subseteq A\}$, respectively. The projection operator 'Proj' is defined as $\text{Proj}_{A}x = \arg\min_{v \in A} \|v - x\|$. and identity map is $id$. A function $\gamma\colon R_{\geq 0} \to R_{\geq 0}$ is a $\mathcal{K}$-function if it is continuous, strictly increasing, and $\gamma(0) = 0$. A $\mathcal{K}$-function $\gamma$ belongs to class $\mathcal{K}_\infty$ if $\gamma(s) \to \infty$ when $s \to \infty$. Finally, a continuous function $\gamma: R_{\geq 0} \times \mathbf{I}_{\geq 0} \to R_{\geq 0}$ belongs to class $\mathcal{KL}$ if, for every fixed $k$, the function $s \mapsto \gamma(s, k)$ is a $\mathcal{K}$-function, and for every fixed $s$, the function $k \mapsto \gamma(s, k)$ is strictly decreasing with $\gamma(s, k) \to 0$ as $k \to \infty$.

\section{PROBLEM DESCRIPTION AND PRELIMINARY}
Consider the uncertain nonlinear system described by
\begin{equation} 
{x_{k + 1}} = f({x_k},{u_k}) + {\delta _k}
\end{equation}
where $ x_k \in R^n $, $ u_k \in R^m $, and $ \delta_k \in R^n $ are the state, control input and disturbance input at time $ k \in \mathbf{I}_{\geq 0} $, respectively, and $ f(x,u) $ is a known function. Assume that the system state is measurable for feedback control and the disturbance input is limited within the compact set $ W \subset R^n $, i.e.,~$ \delta \in W = \left\{ \delta \in R^n : \| \delta \| \le \bar{\delta} \right\}$ with a known upper bound $ \bar{\delta} > 0 $. Moreover, the system (1) is subject to the following constraints for all times:
\begin{equation} 
x_k \in X, \quad u_k \in U, \quad \forall k \in \mathbf{I}_{\geq 0} 
\end{equation}
where $ X \subset R^n $ and $ U \subset R^m $ are compact sets, and both contain the origin in their interior.

\textbf{Assumption 1}. The function $ f(x,u) $ in (1) is Lipschitz continuous on the state for all $ u \in U $, i.e., there exists a Lipschitz constant $ L_f > 0$ such that
\begin{equation} 
\| f(x_1,u) - f(x_2,u) \| \le L_f \, \| x_1 - x_2 \|, \forall x_1, x_2 \in X 
\end{equation}
for all $ u \in U$.

Note that Assumption 1 is not compulsory since it can be replaced with the component-wise Hölder continuity \citep{manzano_componentwise_2021} to reduce the conservatism of controllers.

\textbf{Definition 1} \citep{rawlings_model_2017}. A set $ S \subset R^n $ is called a robust positive invariant (RPI) set of system $ x_{k+1} = F(x_k, \delta_k) $ if, for all \( x_k \in S \) and \( \delta_k \in W \), \( x_{k+1} \in S \).

\medskip

\textbf{Definition 2} \citep{rawlings_model_2017}. A system $
x_{k+1} = F(x_k, \delta_k) $
is said to admit ISS in a RPI set \( S \) if there exist functions \(\alpha \in \mathcal{K}\) and \(\beta \in \mathcal{K}\mathcal{L}\) such that, for all \( x_0 \in S \) and \( \delta \in W \), the system satisfies that
\begin{equation} 
\| x_k \| \le \beta \bigl( \| x_0 \|, k \bigr) + \alpha \Bigl( \sup_{0 \le t \le k} \| \delta_t \| \Bigr), \quad  \forall k \in \mathbf{I}_{\geq 0} . 
\end{equation}
Moreover, the system admits globally ISS if the condition (4) holds for \( S = R^n \).

Consider an unknown real function \( L^{e}_{k}(x,u) \in R \), which measures the economic cost of system (1) at time \( k \ge 0 \). Moreover, the corresponding economic cost function over a time-varying horizon \( N_k > 0 \) is formulated as: 
\begin{equation} 
J_e \bigl(x_k, \mathbf{u}_k, N_k \bigr) = \sum_{i=0}^{N_k - 1} L^{e}_{k}(x_i, u_i)  
\end{equation}
where the control sequence \(\mathbf{u}_k = \{ u_0, u_1, \ldots, u_{N_k-1} \}\). It is noted that the unknown cost \( L^{e}_{k}(x,u) \in R \) may be non-positive definite, non-convex and/or time-varying. Due to the time-varying horizon, the function (5) is also unknown and varying. As a result, the direct minimization of (5) might yield the instability of the resulted closed-loop system \citep{angeli2012average, he2016economic}. 

The goal of the paper is to propose a cost learning-based varying horizon robust EMPC scheme for the system (1), which minimizes the economic cost function (5) taking into account that the economic function is unknown and varying. 

The unknown and varying economic function also causes the unavailable satisfaction of the dissipative conditions on the economical steady state of system (1). The challenge of the addressed goal is then to design the robust EMPC that guarantees recursive feasibility and ISS with respect to the bounded (persistent) disturbance for the unknown and varying economic function (5) while compromising the online computational burden and economic performance of the EMPC.

\textbf{Remark 1.} In general, the economically optimal steady state is calculated by the minimization of the cost \( L^{e}_{k}(x,u) \) provided that it is known and given a prior \citep{ellis_finite-time_2014}. In practice, however, there exist some cases that the economical steady state is pre-specified but the cost is unknown, such as economic cruise control \citep{luo2022multiobjective}, HVAC temperature control \citep{patel2018applications}, etc. In this work, an arbitrary user-specified economic equilibrium point is considered for the EMPC of system (1). Without loss of generality, we assume that the origin is the user-specified equilibrium point of system (1); otherwise, coordinate transformations can be applied. Moreover, to capture the unknown economic cost \( L^{e}_{k}(x,u) \), any regressor is adopted to learn \( L^{e}_{k}(x,u) \) and then the multi-objective control framework \citep{he_stability_2015} integrating a varying horizon adjustment mechanism is used to design the EMPC of system (1).

\section{COST LEARNING-BASED ROBUST EMPC WITHOUT TERMINAL CONSTRAINTS}
\subsection{Regressor-Based Cost Learning}
For the sake of clarity, let $w = [x^{\top}, u^{\top}]^{\top}$. In principle, regressors can be trained based on the data set or the universal alternative assumption \citep{bishop_pattern_2006, murphy_probabilistic_2022}.

\textbf{Assumption 2.} For the unknown economic cost \( L^{e}_{k}(x,u) \) with measured \( x \) and \( u \), there exists a distribution capturing the cost in the form \( y = L^{e}_{k}(w) + \varpi \), for all \( w \in R^{n+m} \), where \( \varpi \) is a random variable.

Note that the distribution in Assumption 2 is unknown and often time-varying. From Assumption 2, one can obtain training data by the stimulation of physical systems.

Consider a general regressor architecture:
\begin{equation} 
\hat{y} = \hat{L}^{e}_{k}(w, \theta) 
\end{equation}
where \(\theta\) is the parameter to be trained, e.g., bias and weights in networks \citep{xiong_adaptive_2023}, and \(\hat{y}\) is the output of the regressor. In general, the output \(\hat{y} \neq y\) due to finite data, random variable \(\varpi\), and regression error. Once the system (1) measures new data, the parameter is updated by a map (e.g., recursive least squares and Bayesian inference), depending on specific regressors \citep{murphy_probabilistic_2022}. Here a specified mixed-kernel map is used to capture the unknown and/or time-varying economic cost.

Let \(\phi_{ij}\) be the \(j\)th kernel function in the \(i\)th component and \({\phi _i} = \sum\limits_{j = 1}^p {{\phi _{ij}}}\) where \(p\) is the kernel number. Note that these kernel functions can be chosen as, e.g., Gaussian, sigmoid kernel, etc. Then the regressor is constructed as
\begin{equation} 
\hat{y} = \theta_k^{\top} \cdot \phi(w) = \sum_{i=1}^{S} \theta_{i,k} \sum_{j=1}^{p} \phi_{ij}(w)
\end{equation}
where \(\phi = [\phi_1, \phi_2, \ldots, \phi_S]^{\top}\), \(S\) is the number of mixed components, \(\theta_{i,k}\) is the weight of the kernel functions. Mixed kernels can enhance the approximation capability of the regressor but increase its complexity \citep{gonen2011multiple}. 

At time $k$, using the new data \((w_{k-1}, y_{k-1})\) and stochastic gradient descent principle, the updating map has the following explicit expression \citep[Chapter 3]{bishop_pattern_2006}:
\begin{equation} 
\begin{aligned}
\theta_k^{\top} & = \gamma_k \cdot (\hat{y}_{k-1} - y_{k-1}) \partial \hat{y}_{k-1}/\partial \theta_k^{\top}\\
& = \gamma_k \cdot \phi(w) \cdot (\hat{y}_{k-1} - y_{k-1})
\end{aligned}
\end{equation} 
where hyperparameter sequence \(\gamma_k\) is the learning rate. Therefore, once the regressor (7) is trained well, the unknown economic cost \(L^{e}_{k}\) can be indirectly optimized using (7) at each time. Note that as \(\hat{L}^{e}_{k}\) is often more complex than \(L^{e}_{k}\), the computational burden of the EMPC will increase accordingly. To this end, in the following, we present a varying horizon EMPC algorithm without terminal constraints for the system (1).

\subsection{Varying Horizon EMPC Algorithm without Terminal sets}  
Consider the system (1) and collect its data \(w_k\) and \(y_k\) in  real time to update the parameters. The approximate economic cost function over the horizon \(N_k\) is then defined as
\begin{equation}
\hat{J}_e\bigl(x_k, \mathbf{u}_k, N_k \bigr) = \sum_{i=0}^{N_k - 1} \hat{L}^{e}_{k}\bigl(x_{i|k}, u_{i|k}, \theta_k \bigr).
\end{equation} 
Due to the time-varying and non-dissipative features of \(\hat{L}^{e}_{k}\), we pick an auxiliary stable cost function
{\small \begin{equation}
J_a\bigl(x_k, \mathbf{u}_k, N_k \bigr) = \sum_{i=0}^{N_k - 1} L_a\bigl(x_{i|k}, u_{i|k}\bigr) + \lambda E_a\bigl(x_{N_k|k}\bigr)
\end{equation} } 
where the positive definite stage cost \(L_a(x,u)\) satisfies
\begin{equation}
L_a(x,u) \ge \gamma_0 \bigl(\| x \|\bigr)
\end{equation} 
for a function \(\gamma_0 \in \mathcal{K}_\infty\), and \(\lambda \ge 1\) is the weight of positive definite penalty function \(E_a(x) \ge 0\) \citep{limon_stability_2006}.

At time \(k \ge 0\) with the state \(x_k \in X\), the FHEOC problem for system (1) is formulated as
\begin{subequations} 
\begin{equation} 
V_e\bigl(x_k, N_k\bigr) = \min_{\mathbf{u}_k}\,\hat{J}_e\bigl(x_k, \mathbf{u}_k, N_k \bigr)
\end{equation} 
\begin{equation} 
\text{s.t.} \quad x_{i+1|k} = f\bigl(x_{i|k}, u_{i|k}\bigr), \quad x_{0|k} = x_k
\end{equation} 
\begin{equation} 
x_{i+1|k} \in X_i, \quad u_{i|k} \in U, \quad \forall i \in \mathbf{I}_1^{N_k-1}
\end{equation} 
\begin{equation} 
\begin{aligned}
J_a\bigl(x_k, \mathbf{u}_k, N_k \bigr) \le N_k d + \lambda a \\
\text{if } N_k < N_{k-1} \text{ or } k=0
\end{aligned} \tag{12d-1} 
\end{equation} 
\begin{equation} 
J_a\bigl(x_k, \mathbf{u}_k, N_k \bigr) \le \Pi\bigl(\lambda, N_k \bigr), \quad \text{if } N_k \ge N_{k-1} \tag{12d-2} 
\end{equation} 
\end{subequations} 
where \(V_e(x, N)\) is the optimal value function of the cost associated with the optimal solution \(\mathbf{u}_k^* = \{u_0^*, u_1^*, \ldots, u_{N_k -1}^*\}\) and the corresponding state sequence \(\mathbf{x}_k^* = \{x_0^*, x_1^*, \ldots, x_{N_k}^*\}\),  \(x_{0|k} = x_k\) is the initial condition, \(X_i \subset X\) is the tightened constraint ensuring the feasibility of the problem. The constraint (12d) is the contractive constraint, the parameters $a$ and $d$ are offline calculated and \(\Pi(x_k, N_k)\) is the contractive sequence on special candidate solutions. Note that only one condition of (12d) is activated at each time depending on the changing horizon. In (12c), the set \(X_i\) is offline reckoned as \citep{limon2002}:
\begin{equation} 
X_i = X \ominus B_i
\end{equation} 
where the ball \(B_i = \{ x \in X : \| x \| \le (L_f^{i} -1)\,\bar{\delta}/(L_f -1) \}\) and \(L_f^{i}\) is the \(i\)th power of constant \(L_f\) for \(i \in \mathbf{I}_1^{N_k-1}\).

\textbf{Assumption 3.}
For the system (1) with \(L_a(x,u)\) and \(E_a(x)\), there exist a number \(a_p >0\) and a control law \(u=\pi(x)\) such that
\begin{equation} 
E_a\bigl(f\bigl(x, \pi(x)\bigr)\bigr) - E_a(x) \le - L_a\bigl(x,\pi(x)\bigr)
\end{equation}
for any \(x \in X_p = \{x \in X : E_a(x)\le a_p \}\). Moreover, there exist scalars \(c_L>0\) and \(c_E>0\) such that
\begin{subequations} 
\begin{equation}
\begin{aligned}
\left| L_a(x_1, u) - L_a(x_2,u) \right| \le c_L \|x_1 - x_2\|   \\
\quad \forall x_1, x_2 \in X,\, u\in U
\end{aligned} 
\end{equation} 
\begin{equation}
\begin{aligned}
\left| E_a(x_1) - E_a(x_2) \right| \le c_E \|x_1 - x_2\|, \forall x_1, x_2 \in X_p\ .
\end{aligned} 
\end{equation} 
\end{subequations} 
\textbf{Remark 2.} Assumption 3 is widely used in MPC \citep{limon2002,sun_robust_2019}. The functions \(L_a(x,u)\) and \(E_a(x)\) are typically chosen as quadratic ones \citep{xiong2025learning}. Then many procedures have been proposed to design \(\pi(x)\) and related set \(X_p\) in Assumption 3, e.g., Jacobian linearization \citep{rawlings_model_2017} or linear matrix inequality \citep{lazar2018computation}.

To determine the parameters \(a\) and \(d\), and remove the traditional terminal constraint during online solving the problem (12), we define the implicit terminal set $ X_T = \{ x \in X : E_a(x)\le a \} $
as the one-step reachable set of \(X_p\) under \(\pi(x)\), i.e.,
\begin{equation} 
f\bigl(x,\pi(x)\bigr) \in X_T, \quad \forall x \in X_p
\end{equation} 
with \(0<a\le a_p\). Let \(a\) be the number in (12d-1), which always exists since \(a_p >0\) under Assumption 3. Moreover, as \(L_a(x,u)\) is positive definite and \(X_T\) contains the origin in its interior, there exists a constant \(d>0\) such that, for all \(x \in X \),
\begin{equation} 
L_a(x,u) > d, \quad \forall x \notin X_T, \quad \forall u \in U.
\end{equation} 
To determine \(N_{k+1}\) and \(\Pi(x_{k+1}, N_{k+1})\) at time \(k+1\), let us consider the following lemma.

\textbf{Lemma 1.}
Under Assumption 1, there exists at least one prediction time \(i \in \mathbf{I}_0^{N_k}\) such that the predicted states satisfy 
\begin{equation} 
\exists i \in \mathbf{I}_0^{N_k} : \quad x^{*}_{i|k}\in X_T
\end{equation}  
provided that the constraint (12d-1) is satisfied at time \(k\).

\textbf{Proof.}  It can be directly proved by contradiction. \(\square\)

From Lemma 1, one cannot specify which predicted state is in \(X_T\), even if (12d-1) is satisfied at all times. Hence, two cases have to be considered as follows.

\textit{Case 1).} $ x_k = x^{*}_{0|k} \in X_T$ and $ x^{*}_{i|k} \not\in X_T, \forall i \in \mathbf{I}_1^{N_k}$. In this case, the system (1) is controlled by the terminal control law \(\pi(x)\), i.e.,
\begin{equation} 
x_{k+1} = f\bigl(x_k,\pi(x_k)\bigr) + \delta_k.
\end{equation} 
\textit{Case 2)}. $\exists i \in \mathbf{I}_1^{N_k} :x^{*}_{i|k} \in X_T$. In this case, the system (1) is controlled by the first element of \(\mathbf{u}_k^{*}\), i.e.,
\begin{equation} 
x_{k+1} = f\bigl(x_k, u_{0|k}^{*}\bigr) + \delta_k.
\end{equation} 
Consider the time when the optimal state first reaches $X_T$, i.e.,
\begin{equation}
t_k^* = \min_{i} \left\{ \exists i \in \mathbf{I}_1^{N_k} : x_{i|k}^* \in X_T \right\}.
\end{equation}
Then for any possible $N_{k+1} \geq t_k^*$, we construct an input candidate sequence
{\small
\begin{equation}
\begin{aligned}
& \bar{\mathbf{u}}(N_{k+1}) 
=  \\ 
& \left\{
\begin{array}{ll}
\left\{
\begin{array}{l}
\pi(\bar{x}_{0|k+1}), \pi(\bar{x}_{1|k+1}), \dots, \\
\pi(\bar{x}_{N_{k+1}|k+1})
\end{array}
\right\} & \text{Case 1)} \\[1ex]
\left\{
\begin{array}{l}
u^*_{1|k}, \dots, u^*_{t_{N_{k+1}}^* -1|k}, \pi(\bar{x}_{t_{N_{k+1}}^*|k+1}), \\
 \dots, \pi(\bar{x}_{N_{k+1}-1|k+1})
\end{array}
\right\} & \text{Case 2)}
\end{array}
\right.
\end{aligned}
\end{equation}}
with the related parameter
\begin{equation}
t_{N_{k+1}}^* = \max_{i} \left\{ i \in \mathbf{I}_{t_k^*} ^{\min\{N_k, N_{k+1}\}} : x_{i|k}^* \in X_T \right\}
\end{equation}
and state sequence
\begin{equation}
\bar{\mathbf{x}}_{k+1} = \left\{ x_{k+1}, \bar{x}_{1|k+1}, \cdots, \bar{x}_{N_{k+1}|k+1} \right\}.
\end{equation}
If the sequence (22) satisfies the constraints (12b) and (12c) at time $k+1$, then $\Pi(x_{k+1}, N_{k+1})$ can be computed as
\begin{equation}
\begin{aligned}
\Pi(x_{k+1}, N_{k+1}) & = (1 - \mu)\left(V^e_a(x_k, N_k) + \bar{\delta} \xi_{N_k} \right) \\
& + \mu J_a(x_k, \bar{\mathbf{u}}(N_{k+1}), N_{k+1})
\end{aligned}
\end{equation} 
where the compromise factor $\mu \in [0,1]$ balances the stability and economic performance of EMPC, the auxiliary value function
\begin{equation}
V^e_a(x_k, N_k) = J_a(x_k, \mathbf{u}^*_k, N_k)
\end{equation}
and the number
\begin{equation}
{\xi _{{N_k}}} = \lambda {c_E}{L_f^{{N_k} - 1}} + {c_L}{{(L_f^{{N_k} - 1} - 1)} \mathord{\left/
 {\vphantom {{(L_f^{{N_k} - 1} - 1)} {({L_f} - 1)}}} \right.
 \kern-\nulldelimiterspace} {({L_f} - 1)}}
\end{equation}
For the system (1), we introduce two sets
\begin{align}
\Gamma = \Big\{ N \in \mathbf{I}_{\geq 1} :\ & c_E L_f^{i-1} \bar{\delta} \leq a_p - a, \notag \\
& X_p \subseteq X_{N-1},\ \forall i \in \mathbf{I}_1^N \Big\}
\end{align}
and for the compromise factor $\mu \in [0,1]$,
\begin{equation}
\Theta = \left\{ N \in \mathbf{I}_{\geq 1} : \alpha_2 \left( (\mu \gamma_0)^{-1} \bar{\delta} \xi_{N} \right) \leq Nd+ \lambda a \right\}. 
\end{equation}
where function  $\alpha_2$ will be defined in Assumption 4 later. 

Moreover, we define
\begin{equation}
{\tilde N_{k + 1}} = \max \left\{ {\left\lceil {t_k^{\text{*}}{\upsilon _k} + (1 - {\upsilon _k}){N_k}} \right\rceil  + {\sigma _k},1} \right\}
\end{equation}
where the numbers $\upsilon _k \in [0,1]$ and ${{\sigma _k}} \geqslant 0$ tune the horizon increasing and decreasing, respectively. The following iterative algorithm 1 is used to compute the actual horizon $N_{k+1}$ of (12).

\begin{table}[ht]
\centering
\begin{tabular}{p{8cm}}
\toprule
\textbf{Algorithm 1:} Iteratively calculating $N_{k+1}$ of EMPC  \\
\midrule
Given the predicted sequence $\mathbf{x}_k^*$, judge the case of 1) or 2) occurs and measure $x_{k+1}$. Compute $t^*_k$ in (21) and $\tilde{N}_{k+1}$ in (30); let $\Xi_{k+1} = \emptyset$. \\
\textbf{For} $i = t^*_k, \dots, \max\{\tilde{N}_{k+1}, N_k\}$ \\ 
\hspace{1em}  Construct $\bar{\mathbf{u}}(i) $ in (22) and calculate $J_a(x_{k+1}, \bar{\mathbf{u}}(i) , i)$ \\
\hspace{2em} \textbf{If} $J_a(x_{k+1}, \bar{\mathbf{u}}(i) , i) \leq id + \lambda a$ \\  
\hspace{3em}  $\Xi_{k+1} \gets \Xi_{k+1} \cup \{i\}$ \\
 \textbf{End} \\
 \textbf{Return} $N_{k+1} = \text{Proj}_{\Gamma \cap \Theta \cap \Xi_{k+1}} \tilde{N}_{k+1}$ \\
\bottomrule
\end{tabular}
\end{table}

It is noticed that no optimization is involved in Algorithm 1 as $\bar{\mathbf{u}}(i)$ and $J_a(x_{k+1}, \bar{\mathbf{u}}(i), i)$ are explicitly evaluated by (22) and (10), respectively. Hence, the extra computational burden of the varying horizon robust EMPC is slight and can be negligible. Moreover, by tuning  $\upsilon _k $  and $\sigma_k$, the rapid  reduction of the horizon can facilitate fast optimization of the FHEOC problem in (12). In the following, the new robust EMPC algorithm is presented as Algorithm~2.

\begin{table}[ht] 
\centering
\begin{tabular}{p{8cm}}
\toprule
\textbf{Algorithm 2:} Varying horizon robust EMPC algorithm \\
\midrule
\textbf{OFFLINE OPERATION} \\
  \hangindent=2em
  \hangafter=1
\textbf{Step 1.} (Controller Initialization) Let $k = 0$; Given the steady-state point $(x_{s}, u_{s})$ and choose $L_a$, $E_{a}$, $\lambda$, $\upsilon _k$, $\sigma_k$. Calculate $a$, $d$, $N_0$ and sets $\Theta$ and $\Gamma$ such that $N_0 \in \Theta \cap \Gamma$ and (12d-1) is satisfied. \\
  \hangindent=2em
  \hangafter=1
\textbf{Step 2.} (Regressor Initialization) Choose the architecture (7), specify the parameter refresh map and randomly initialize $\theta_0$. \\
\addlinespace
\textbf{ONLINE OPERATION} \\
  \hangindent=2em
  \hangafter=1
\textbf{Step 3.} Measure $x_k$ and solve the FHEOC problem (12) to derive $\mathbf{u}_k^*$. \\
  \hangindent=2em
  \hangafter=1
\textbf{Step 4.} If Case 1) occurs, the control law $u_k = \pi(x_k)$ is applied to the system (1); Otherwise, the first element of $\mathbf{u}_k^*$ is applied to (1). \\
  \hangindent=2em
  \hangafter=1
\textbf{Step 5.} Collect the real-time data $(w_{k-1}, y_{k-1})$ and update parameter $\theta_k$ using (8). \\
  \hangindent=2em
  \hangafter=1
\textbf{Step 6.} Calculate $\tilde{N}_{k+1}$ by (30) and determine $N_{k+1}$ by executing Algorithm 1; Set $k = k + 1$ and return to Step 3. \\
\bottomrule
\end{tabular}
\end{table}

\section{FEASIBILITY AND ROBUST STABILITY}  
Consider the system (1) and FHEOC problem (12). We first analyze the feasibility of the problem and then derive the robust stability results for the closed-loop system in (19) or (20) with respect to the disturbance $\delta$. 

\textbf{Lemma 2.} Suppose that the problem (12) with (12d-1) is feasible at time $k$. Then under Assumptions 1 and 3, the control sequence (22) for all $ N_{k+1} \geq N_k$ satisfies the constraints (12b) and (12c) at time $k+1$ if the set (28) is nonempty and $N_k \in \Gamma$. Furthermore, there exists the following inequality:
\begin{equation}
\begin{aligned}
& J_a(x_{k+1}, \bar{\mathbf{u}}(N_{k+1}), N_{k+1}) \\
\leq & V_a^e(x_k, N_k) - \mu L_a(x_k, u^*_{0|k}) + \bar{\delta} \xi_{N_k}.
\end{aligned}
\end{equation}
\textbf{Proof.} See Appendix.  $\Box$

In (28), the inequality condition characterizes the coupling between the horizon and disturbance bound, ensuring recursive feasibility for decreasing horizons, and the other one guarantees that the nominal invariant set $X_p$ remains strictly be contained within the tightened constraints~\citep{limon2002}. Moreover, from the positive definiteness of $L_a(x,u)$, Eq.~(25) and (31), one can obtain that
$ J_a(x_{k+1}, \bar{\mathbf{u}}(N_{k+1}), N_{k+1}) \leq \Pi(x_{k+1}, N_{k+1}).$
Therefore, the sequence (22) is feasible for any $N_{k+1} \geq N_k$ at $k+1$. Nevertheless, feasibility at $k+1$ does not guarantee feasibility at all future times as the conditions (12d-1) and $N \in \Gamma$ may not be satisfied thereafter. To address this problem, a new assumption is required.

\textbf{Assumption 4.} For the auxiliary value function $V_a^e(x, N)$, there exists a function $\alpha_2 \in \mathcal{K}_\infty$ such that
\begin{equation}
V_a^e(x, N) \leq \alpha_2(\|x\|), \quad \forall N \in \Gamma
\end{equation}
and $id - \mu \gamma_0 \alpha_2^{-1}$ is a $\mathcal{K}$-function. Moreover, there exists a lower bound $\mu_{\min} \in (0,1]$ such that the set $\Theta \cap \Gamma \neq \emptyset$ for $\mu \in [\mu_{\min}, 1]$.

Notably, the inequality (32) is the so-called controllability condition widely used in EMPC~\citep{faulwasser2018economic,rawlings_model_2017}. Also, the nonempty intersection is unrestrictive if $\bar{\delta}$ is small enough because
\begin{subequations}
\begin{equation}
X_p \subseteq X = \lim_{\bar{\delta} \to 0} X_{N-1}, \quad \forall N \in \mathbf{I}_{\geq 1}
\end{equation}
\begin{equation}
\lim_{\bar{\delta} \to 0} \alpha_2 \left( (\mu \gamma_0)^{-1} \bar{\delta} \xi_N \right) = 0 \leq Nd + \lambda a, \forall N \in \mathbf{I}_{\geq 1}. 
\end{equation}
\end{subequations}
Moreover, the $\mathcal{K}_\infty$-function $\alpha_2$ may not be easy to find for a learned economic cost. To this end, the two-stage strategy~\citep{xiong2024two} can be employed to ensure the reliability of the function in an origin neighborhood. Then, by combining this with the boundedness of the cost function and compactness of sets $X$ and $U$, one can derive an appropriate function $\alpha_2$.

\textbf{Lemma 3.} Consider the problem (12) with $N_k \in \Theta \cap \Gamma$. If the constraint (12d-1) is satisfied at time $k$ and Assumptions 1, 3 and 4 hold, then the horizon $N_{k+1}$ obtained by Algorithm~1 is well-defined and the control sequence (22) satisfies that
\begin{equation}
J_a(x_{k+1}, \bar{\mathbf{u}}(N_{k+1}), N_{k+1}) \leq N_{k+1} d + \lambda a.
\end{equation}
\textbf{Proof.} To show the existence of at least one well-defined $N_{k+1}$, it suffices to ensure that $N_{k+1} = N_k$ under the sequence (22) satisfies the required conditions. To this end, we define the set
{\small 
\begin{equation}
B_{N_k}^{\mu} = \left\{ x \in X : V_a^e(x_k, N_k) \leq \alpha_2 \left( (\mu \gamma_0)^{-1} \bar{\delta} \xi_{N_k} \right) \right\}.
\end{equation}}
At time step $k+1$, we consider the following two parts:

\textbf{Part 1.} If $x_{k} \in B_{N_k}^{\mu}$, then from Lemma 2, there is
{\small \begin{equation}
\begin{aligned}
  & {J_a}({x_{k + 1}},{{\mathbf{\bar u}}}({N_{k + 1}}),{N_{k + 1}}) \hfill \\
  &\mathop  \leqslant \limits^{(31)} V_a^e({x_k},{N_k}) - \mu {L_a}({x_k},u_{0|k}^*) + \bar \delta {\xi _{{N_k}}} \hfill \\
  &\mathop  \leqslant \limits^{(11),(35)} V_a^e({x_k},{N_k}) - \mu {\gamma _0}\left( {\alpha _2^{ - 1}(V_a^e({x_k},{N_k})} \right) + \bar \delta {\xi _{{N_k}}} \hfill \\
  &\mathop  \leqslant {\alpha _2}\left( {{{(\mu {\gamma _0})}^{ - 1}}\bar \delta {\xi _{{N_k}}}} \right) - \bar \delta {\xi _{{N_k}}} + \bar \delta {\xi _{{N_k}}} \hfill \\
  &\mathop  \leqslant \limits^{(29)} {N_k}d + \lambda a \hfill \\ 
\end{aligned}
\end{equation}}
where the third inequality is derived from the fact that $id - \mu \gamma_0 \alpha_2^{-1}$ belongs to class $\mathcal{K}$.

\textbf{Part 2.} If $x_{k+1} \notin B_{N_k}^{\mu}$, then we have
\begin{equation}
\bar{\delta} \xi_{N_k} - \mu \gamma_0 \left( \alpha_2^{-1}(V_a^e(x_k, N_k)) \right) < 0.
\end{equation}
Thus, it is derived that
\begin{equation}
\begin{aligned}
  &{J_a}({x_{k + 1}},{{\mathbf{\bar u}}}({N_{k + 1}}),{N_{k + 1}}) \hfill \\
  &\mathop  \leqslant \limits^{(31)} V_a^e({x_k},{N_k}) - \mu {L_a}({x_k},u_{0|k}^*) + \bar \delta {\xi _{{N_k}}} \hfill \\
  &\leqslant V_a^e({x_k},{N_k}) - \mu {\gamma _0}\left( {\alpha _2^{ - 1}(V_a^e({x_k},{N_k})} \right) + \bar \delta {\xi _{{N_k}}} \hfill \\
  &\leqslant {N_k}d + \lambda a. \hfill \\ 
\end{aligned}
\end{equation}
This completes the proof of Lemma~3.  $\Box$

\begin{figure} 
\begin{center}
\includegraphics[height=4cm]{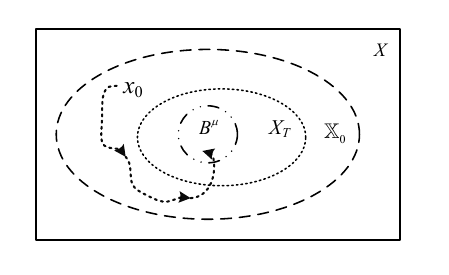}    
\caption{The relationships of the state sets $X$ (solid line), ${\mathbb{X}_0}$ (dashed line), $X_T$ (dotted line) and $B^\mu$ (dash-dotted line). The anticipated closed-loop trajectory (arrowed line) converges to the interior of $B^\mu$ and maintains within it.}
\label{fig2}                                 
\end{center}                                 
\end{figure} 

Let ${\mathbb{X}_0}$ be the initial feasible state set from which the problem (12) is always feasible at time $k=0$. Furthermore, we define the set $B^{\mu}$ as
\begin{equation}
\begin{aligned}
B^{\mu} = \left\{ x \in X : V_a^e(x_k, N_k) \leq \alpha_2 \left( (\mu \gamma_0)^{-1} \bar{\delta} \xi \right) \right\}
\end{aligned}
\end{equation}
with $\xi = \max_{N \in \Theta \cap \Gamma} \xi_N$. The set $B^{\mu}$ is the uniform outer approximation to the sets $B^{\mu}_{N_k}$. To clarify, Fig. 2 illustrates the relations of these state sets and the expected behaviors of the state trajectories, where $X$ is the state constraint, ${\mathbb{X}_0}$ is the initial feasible set, $X_T$ is the terminal set,  $B_N^\mu$ is the RPI set related to fixed horizon $N$, and the arrowed line is the anticipated closed-loop trajectory. 

\textbf{Lemma 4.} For the system (1) and the problem (12) with Assumptions 1–4, the conclusion (18) holds whenever $N_{k+1} \geq N_k$.

\textbf{Proof.} It is observed from the problem (12) that the constraint (12d) is activated from not (12d-2) but (12d-1) at initial time $k=0$. Namely, the constraint (12d-2) is activated later than the one (12d-1). Denote $\bar{k}$ as the latest time at which (12d-1) is activated. Two cases are considered as follows.

If $x_{\bar{k}} \notin B^{\mu}$, we have that
\begin{equation}
\begin{aligned}
 & \Pi ({x_{\bar k + 1}},{N_{\bar k + 1}})\\
 & \mathop  \leqslant \limits^{(25),(31)} V_a^e({x_{\bar k}},{N_{\bar k}}) - \mu {L_a}({x_{\bar k}},u_{0|\bar k}^*) + \bar \delta \xi  \hfill \\
 \; &  \mathop  \leqslant \limits^{(39),(11)} \Pi ({x_{\bar k}},{N_{\bar k}}). \hfill \\
\end{aligned} 
\end{equation} 
Iterating (40), it is obtained that $\Pi(x_{k+1}, N_{k+1}) \leq \Pi(x_{\bar{k}}, N_{\bar{k}})$. Then the non-decreasing horizon under (12d-2) implies that
\begin{equation}
\begin{aligned}
\Pi ({x_{k + 1}},{N_{k + 1}})& \leqslant \Pi ({x_{\bar k}},{N_{\bar k}}) \\
& \mathop \leqslant  {N_{\bar k}}d + \lambda a\mathop  \leqslant {N_{k + 1}}d + \lambda a.
\end{aligned}
\end{equation}
If $x_{\bar{k}} \in B^{\mu}$, similar to (36), one can derive that
{\small 
\begin{equation}
\begin{gathered}
  \Pi ({x_{\bar k + 1}},{N_{\bar k + 1}}) \leqslant V_a^e({x_{\bar k}},{N_{\bar k}}) - \mu {L_a}({x_{\bar k}},u_{0|\bar k}^*) + \bar \delta \xi  \hfill \\
  \quad \quad \quad \quad \quad  \leqslant {\alpha _2}\left( {{{(\mu {\gamma _0})}^{ - 1}}\bar \delta \xi } \right) \hfill \\ 
\end{gathered} 
\end{equation}} 
which implies that $\Pi(x_{k+1}, N_{k+1}) \leq \alpha_2 \left( (\mu \gamma_0)^{-1} \bar{\delta} \xi \right) \leq N_{k+1} d + \lambda a$ for any $k \geq \bar{k}$. Combining two cases, the conclusion of Lemma 4 is proven.  $\Box$

\textbf{Remark 3:} From the conclusions of Lemmas 1 and 4, it is directly obtained that whether activating the constraint (12d-1) or (12d-2), there exists at least one predicted state that can be \textit{implicitly} driven into the terminal set at each time. Then, from the perspective of computation (e.g., \citet{limon_stability_2006}), this will lessen the computational burden of online solving the problem (12).

\textbf{Theorem 1.} Consider the system (1) with ${\mathbb{X}_0}$ and ${N_{\text{0}}} \in \Theta  \cap \Gamma $. Under Assumptions 1-4, the problem (12) is recursively feasible. 

\textbf{Proof.} Assume that the problem (12) is feasible at time $k$ and then the candidate sequence (22) is constructed at $k+1$. Depending on the horizon change, two possibilities are discussed as follows.

1) If $N_{k+1} \geq N_k$, then from Lemma~2, it is known that the candidate (22) satisfies the constraints (12b) and (12c). To verify the satisfaction of (12d-2), we have that
{\small 
\begin{equation}
J_a(x_{k+1}, \bar{\mathbf{u}}(N_{k+1}), N_{k+1}) 
\mathop  \leqslant \limits^{(31)} V_a^e(x_k, N_k) + \bar{\delta} \xi_{N_{k}}
\end{equation} }
which yields
\begin{equation}
\begin{aligned}
& J_a(x_{k+1}, \bar{\mathbf{u}}(N_{k+1}), N_{k+1})  \\
&\leq (1 - \mu)\left( V_a^e(x_k, N_k) + \bar{\delta} \xi_{N_{k}} \right)\\
&\quad + \mu J_a(x_{k+1}, \bar{\mathbf{u}}(N_{k+1}), N_{k+1})\\
&\leq \Pi(x_{k+1}, N_{k+1}). 
\end{aligned}
\end{equation}
2) If $N_{k+1} < N_k$, the satisfaction of (12b) and (12c) in Case 1) is trivial, and one can derive that
\begin{equation}
J_a(x_{k+1}, \bar{\mathbf{u}}(N_{k+1}), N_{k+1}) \leq \lambda a
\end{equation}
which implies that (12d-1) is fulfilled directly. For Case 2) at $k+1$, the consequence (22), tightened constraints and Assumption 3 ensure that
\begin{equation}
\bar{x}_{i-1|k+1} \in X \ominus B_i \oplus W \subseteq X_{i-1}, \quad \forall i \in \mathbf{I}_1^{t^*_{N_{k+1}}}.
\end{equation}
Moreover, from Algorithm 1 in Case 2), it is derived that $N_{k+1} \in \Xi_{k+1}$. As a result, we have that
\begin{equation}
J_a(x_{k+1}, \bar{\mathbf{u}}(N_{k+1}), N_{k+1}) \leq N_{k+1} d + \lambda a
\end{equation}
which ensures that (12d-1) is satisfied. $\Box$

In varying horizon EMPC, the value function \( V_a^e(x, N) \) is not
necessarily an ISS-Lyapunov function of the closed-loop system due to the varying prediction. One method to address this problem is to resort to the one-step reachable set. To this end, we
define the outer approximation of one-step reachable set of \( \mathbb{X}_0 \) as
\begin{equation}
\mathbb{X}_1 = \left\{ x \in X : V_a^e(x,N) \leq r_1 \right\}, \quad \forall N \in \Theta \cap \Gamma
\end{equation}
where \( r_1 = \gamma(\bar{\delta}) + \max_{x \in \mathbb{X}_0} \left\{ V_a^e(x,N_0) - \mu \gamma_0(x) \right\} \). Then, the robust stability result is presented as Theorem~2. 

\textbf{Theorem 2.}
Consider the system (1) with Algorithm~2. If Assumptions~1--4 hold and \( \mathbb{X}_1 \subseteq \mathbb{X}_0 \), \( N_0 \in \Theta \cap \Gamma \), then the system with the EMPC under non-decreasing horizons admits ISS in the RPI set \( \mathbb{X}_0 \).

\textbf{Proof.}
Due to the positive definiteness of \( L_a(x,u) \) and \(E_a(x) \) and Assumption~4, the value function \( V_a^e(x_k, N_k) \) satisfies that
\begin{equation}
\gamma_0(\|x_k\|) \leq V_a^e(x_k, N_k) \leq \alpha_2(\|x_k\|)
\end{equation}
which, together with the proof of Lemma~4 and (12d-2), yields
\begin{equation}
V_a^e(x_k, N_k) \leq V_a^e(x_{k-1}, N_{k-1}) - \mu \gamma_0(\|x_k\|) + \bar{\delta} \xi. 
\end{equation}
Then from Lemma 3.5 in \cite{jiang_input--state_2001}, there exists a function \( \hat{\beta} \in \mathcal{KL} \) such that the closed-loop system satisfies that
{\small
\begin{equation}
\left\| {{x_k}} \right\| \leqslant \gamma _0^{ - 1}\left( {\hat \beta \left( {{\alpha _{\text{2}}}\left( {\left\| {{x_{\text{0}}}} \right\|} \right),k} \right)} \right) + \gamma _0^{ - 1}\left( {{\alpha _{\text{2}}}\left( {{{(\mu {\gamma _0})}^{ - 1}}\bar \delta \xi } \right)} \right). 
\end{equation}}
Further consider any \( x_0 \in \mathbb{X}_0 \) with \( N_0 \in \Theta \cap \Gamma \). From (48) and (50), it is obtained that \( V_a^e(x_1, N_1) \leq r_1 \). Then we have that \( x_1 \in \mathbb{X}_1 \subseteq \mathbb{X}_0 \),
which implies that \( \mathbb{X}_0 \) is a RPI set under non-decreasing horizons.
Therefore, the closed-loop system under non-decreasing horizons has ISS in \( \mathbb{X}_0 \). The proof is completed. $\Box$

If the horizon \( N_0 \) is fixed from the initial time up to the time \( s \),
then the condition ${\mathbb{X}_1} \subseteq {\mathbb{X}_0}$ can be relaxed. To do this, we introduce the following recursive computations:
{\small
\begin{subequations}
\begin{equation}
\mathbb{X}_s = \left\{ x \in X : V_a^e(x, N_0) \leq r_s \right\} \subseteq \mathbb{X}_{s-1}, \forall s \in \mathbf{I}_{\geq1} 
\end{equation}
\begin{equation}
{r_s} =  \gamma (\bar \delta )+ \mathop {\max }\limits_{x \in {\mathbb{X}_{s - 1}}} V_a^e(x,{N_0}) - \mu {\gamma _0}(x).  
\end{equation}
\end{subequations}}
Then the condition ${\mathbb{X}_1} \subseteq {\mathbb{X}_0}$ can
be relaxed by the following \( s \)-step condition:
{\small 
\begin{equation}
\mathbb{X}_s = \left\{ x \in X : V_a^e(x, N) \leq r_s \right\} \subseteq \mathbb{X}_0, \quad \forall N \in \Theta \cap \Gamma.
\end{equation}}
Note that there exists at least one time moment \( s > 1 \) such that the
condition (53) holds. Because Algorithm~2 returns to the standard
EMPC with a fixed horizon as \( s \rightarrow \infty \), under which \( V_a^e(x_k, N_k) \) is an ISS-Lyapunov function of the closed-loop system \citep{xiong2025learning}.

It is noticed that the condition (52) involves the optimization of value functions, which typically leads to a bilevel optimization problem being hard to be solved. To circumvent this problem, here we use the Monte Carlo sampling method. Namely, randomly pick $q$ initial points within \( \mathbb{X}_0 \)  and compute each closed-loop trajectory for s steps under horizon $N_0$. Then, the condition (53) can be approximated by 
{\small
\begin{equation}
\left\{ x \in X : V_a^e(x, N) \leq \max_{i = 1, \dots, q} V_a^e(x_s^i, N_0) \approx r_s \right\} \subseteq \mathbb{X}_0.
\end{equation}}
On the other hand, from a practical perspective, it is helpful to move out the condition~(53) since this allows the EMPC to explore the regions beyond \( \mathbb{X}_0 \)
while satisfying the constraint~(2).

\textbf{Remark 4.} The feasibility and stability proofs do not involve complicated dissipative verification or priori Lyapunov functions \citep{ellis_finite-time_2014} while only relying on the classical robust terminal set in Assumption 1 and a small enough error boundary, which, as one of the main advantages of multi-objective EMPC, extends the application scope of the algorithm. In addition, there are no regressor property requirements (e.g., convergence, error boundary, as in \citet{manzano_componentwise_2021}), implying that the proposed MPC algorithm is compatible with all existing regression algorithms, e.g., \citet{murphy_probabilistic_2022, bishop_pattern_2006}. 

\section{EXPERIMENTS}
In this section, two examples are used to verify the effectiveness of the proposed EMPC scheme. The first one is the bilinear CSTR, where some comparative experiments are carried out to illustrate the necessity and advantages of the proposed EMPC. In second example, a four-tank system is utilized to evaluate the performance improvement resulted from cost learning and the influence of varying horizons on the results. All simulation experiments are performed on a Dell Laptop with an Intel 10th-generation i5 processor, utilizing sequential quadratic programming to online solve the involved optimization problems. 

\subsection{A Bilinear CSTR}
Consider a bilinear CSTR system \citep{amrit2011economic} described by 
\begin{equation}
\begin{aligned}
\dot{c}_A &= \frac{Q(c_{Af} - c_A)}{V_R} - k_r c_A + \delta^1 \\
\dot{c}_B &= \frac{Q(c_{Bf} - c_B)}{V_R} + k_r c_A + \delta^2
\end{aligned}
\end{equation}
where $c_{Af}=1$, $c_{Bf}=0$ are molar concentrations of $A$ and $B$ respectively. The volume of the reactor is $V_R=10L/\min$, and rate constant is $k_r=1.2L/(\text{mol}\cdot\text{min})$. The state and control input are selected as $x=(c_A,c_B)$, and $u=Q$ representing the flow
rate. The state and control constraints are $X=[0,1]\times[0,1]$, $U=[0,20]$. Let the disturbance $\delta^1$ and $\delta^2$ are both uniformly distributed over the set $W=[-1\times10^{-4},1\times10^{-4}]$. We choose a sampling period of $T=0.5$ min and discretize the system using the Euler method.

In this study, the multi-objective nominal EMPC (i.e., NEMPC) algorithm \citep{he2016economic} and terminal constraint-based EMPC (i.e., TEMPC) algorithm for disturbed systems \citep{defeng2019input} are selected to compare with our varying horizon EMPC (i.e. VHEMPC). For fair comparisons, the economic cost is assumed known as $L_k^e(x,u)=-c_B=-x_2$ and fix the horizon of $N_0=7$. Notably, the system setting is non-dissipative \citep{amrit2011economic}. The control goal of the CSTR is to find the EMPC that steers the CSTR system from the initial state $(0.40, 0.99)$ to a user-specified steady state $x_s=(0.25,0.75)$ with input $u_s=4$. Moreover, the auxiliary stable cost function is selected as
\begin{equation} 
\begin{aligned}
J_a(x_k, \mathbf{u}_k, N_k) & = \sum_{i=0}^{N_k-1} (\bar{x}_{i|k}^\mathrm{T} \mathbf{Q} \bar{x}_{i|k} + \bar{\mathbf{u}}_{i|k}^\mathrm{T} \mathbf{R} \bar{\mathbf{u}}_{i|k}) \\
& + \bar{x}_{N_k|k}^\mathrm{T} \mathbf{P} \bar{\mathbf{x}}_{N_k|k}  
\end{aligned}
\end{equation} 
where $\bar{x} = x - x_s$ and $\bar{u} = u - u_s$, which transfer the equilibrium point to the origin, and weights $\mathbf{R}=0.1$ and $\mathbf{Q}=\text{diag}\{1,1\}$. Then the terminal penalty matrix is computed by the function `dlqr' of MatLab, i.e.,
$\mathbf{P} = \begin{bmatrix}
2.4565 & 1.5597 \\
1.5597 & 2.7437
\end{bmatrix}.$
The terminal set parameters are further estimated using the method in \citet{manzano2019output}, i.e., $\alpha_p=0.097, \alpha=0.0705$, and $d=0.0078$. Moreover, we set $\mu = 0.95$ and estimate the parameters $c_L = 2.1184$, $c_E = 0.8612$ and $L_f=1.1854$.
\begin{figure}
\begin{center}
\includegraphics[width=0.5\textwidth]{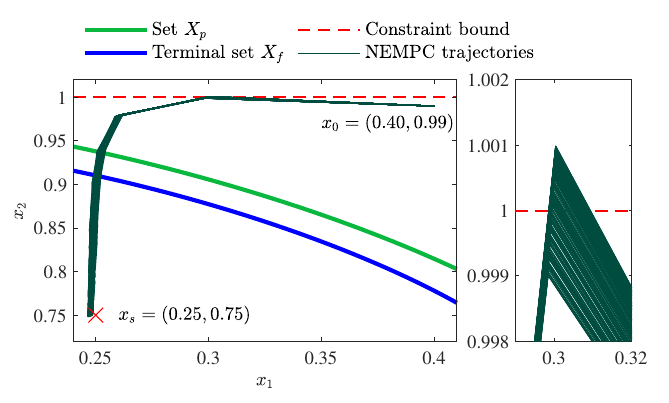}    
\caption{Mento Carlo simulation results on NEMPC} 
\label{fig3}                                 
\end{center}                                 
\end{figure}
\begin{figure}
\begin{center}
\includegraphics[width=0.5\textwidth]{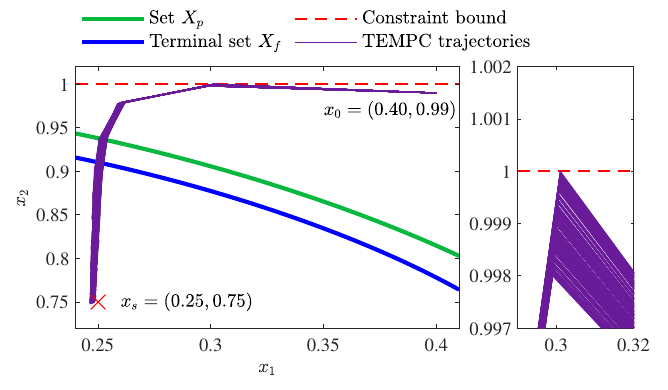}    
\caption{Mento Carlo simulation results on TEMPC} 
\label{fig4}                                 
\end{center}                                 
\end{figure}
\begin{figure}
\begin{center}
\includegraphics[width=0.5\textwidth]{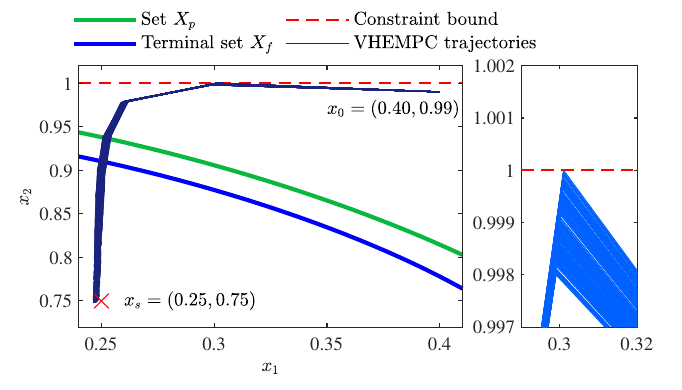}    
\caption{Mento Carlo simulation results on VHEMPC} 
\label{fig5}                                 
\end{center}                                 
\end{figure}
\begin{figure}
\begin{center}
\includegraphics[width=0.45\textwidth]{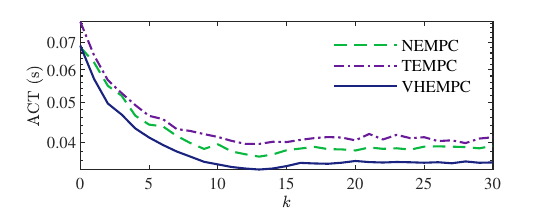}    
\caption{ACT of three algorithms } 
\label{fig1}                                 
\end{center}                                 
\end{figure}
In simulation, 200 Monte Carlo simulation experiments are conducted to evaluate the performance of three EMPC algorithms, with each simulation running for 30 steps. The results are shown in Figs. 3–6, where Figs. 3-5 separately show the results on NEMPC, TEMPC and VHEMPC and Fig. 6 pictures the average computation time (ACT) of online solving the optimization problems in three algorithms at each time. As illustrated in Figs. 3–5, the state trajectories driven by the three EMPC converge to a small neighbourhood of the predefined steady state. However, when the trajectory approaches the boundary of the constraint set, both VHEMPC and TEMPC consistently enforce constraint satisfaction by appropriately tightening the constraint set $X$. In contrast, since NEMPC directly applies the nominal controller, there exists a constraint violation rate of 50\%.

Moreover, Fig. 6 shows the ACT of the three algorithms per time moment, where the dash-dotted, dash and solid lines are associated with NEMPC, TEMPC and VHEMPC, respectively. From Fig. 6, it is observed that VHEMPC has the minimum computational time than the other two algorithms. The main reason is the removal of terminal constraint in VHEMPC whilst the other two algorithms are subject to the explicit terminal constraints. Again, NEMPC exhibits slightly better computation efficiency than TEMPC, owing to the absence of online constraint-tightening procedures.

Additionally, we calculate the average economic costs over the time interval $[0,10]$ as $-0.8980$, $-0.8978$, and $-0.8978$ for NEMPC, TEMPC, and VHEMPC, respectively. While NEMPC has a better transient economic performance than the other two algorithms, it achieves this performance at the expense of frequent constraint violations, particularly at time $k=1$.

In summary, the proposed VHEMPC significantly reduces computational burden while preserving robustness and achieving comparable performance to the terminal-constraint-based scheme.

\subsection{Four-Tank System}

\begin{figure}
\begin{center}
\includegraphics[height=9.2cm]{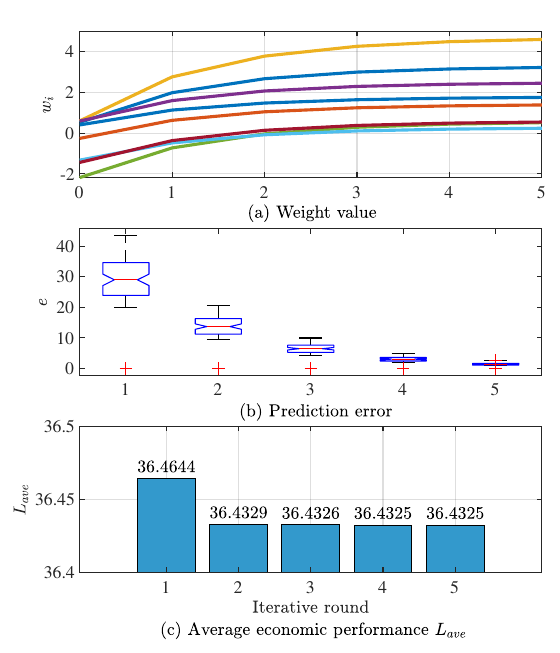}    
\caption{Learning effect and control performance under online approximator. (a) Weight changes under random initialization; (b) Test errors; (c) Average economic performance of the closed-loop trajectories.} 
\label{fig1}                                 
\end{center}                                 
\end{figure}

To further evaluate the effects of cost learning and varying horizons on the performance and computational burden of the proposed VHEMPC, a more complex four-tank system is used in this subsection. The system consists of four diagonally interconnected tanks that regulate flow to maintain the levels of water in tanks. According to \citet{djorge2018robust}, the dynamics of the tank system is described by the differential equations
\begin{equation}
\begin{split}
\frac{dh_1}{dt} &= -\frac{a_1}{S} \sqrt{2gh_1} + \frac{a_3}{S} \sqrt{2gh_3} + \frac{\gamma_\alpha}{S} \cdot \frac{q_a}{3600} \\
\frac{dh_2}{dt} &= -\frac{a_2}{S} \sqrt{2gh_2} + \frac{a_4}{S} \sqrt{2gh_4} + \frac{\gamma_\alpha}{S} \cdot \frac{q_b}{3600} \\
\frac{dh_3}{dt} &= -\frac{a_1}{S} \sqrt{2gh_3} + \frac{1 - \gamma_\alpha}{S} \cdot \frac{q_b}{3600} \\
\frac{dh_4}{dt} &= -\frac{a_4}{S} \sqrt{2gh_4} + \frac{1 - \gamma_\alpha}{S} \cdot \frac{q_a}{3600} \\
\end{split}
\end{equation}
where $h_i$ and $a_i$ are the water level and the cross section of the outlet hole of tank $i = 1,2,3,4$, respectively, $q_a$ and $q_b$ are the pump flows, $S$ is the cross section of the tanks, $\gamma_\alpha$ is the flow parameter of the tanks and $g$ is the gravity constant.

\begin{figure*}[t]
\begin{center}
\includegraphics[height=10.2cm]{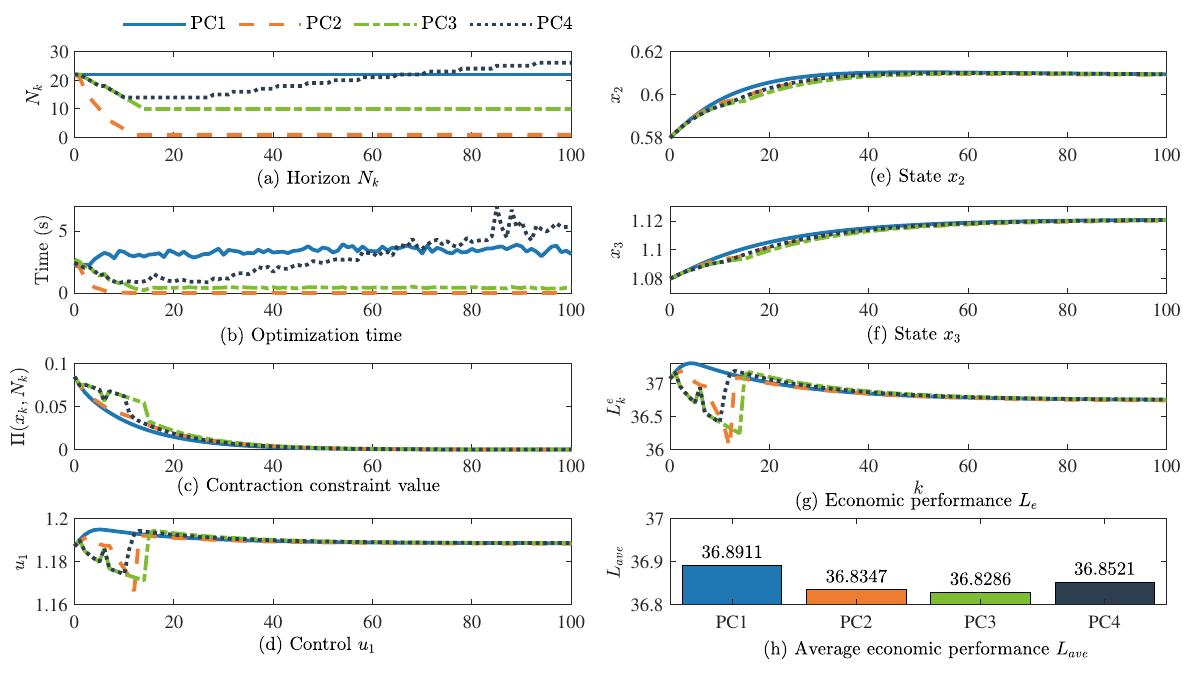}    
\caption{Four-tank experiment results under four different parameters. (a) Horizon changes; (b) Online optimization time of economic FHEOC; (c) Contraction constraint value of economic FHEOC; (d) Control input $u_1$; (e) Second state $x_2$ of the closed-loop system; (f) Third state $x_3$ of the closed-loop system; (g) Real stage economic performance of the closed-loop system; (h) Average economic performance.}    
\label{fig2}                                 
\end{center}                                 
\end{figure*}

In this study, the parameters of system (55) are adopted from \citet{djorge2018robust}, i.e., $S = 0.4$, $a_1 = 1.310 \times 10^{-4}$, $a_2 = 1.507 \times 10^{-4}$, $a_3$=$9.267 \times 10^{-4}$, $a_4$=$9.816 \times 10^{-5}$, and $\gamma_\alpha$=$0.3$, $\gamma_b$=$0.4$. 
Let the sampling time be $10$ s. The state and control input of the tank system are defined as $x = [h_1, h_2, h_3, h_4]^\mathrm{T}$ and $u = [q_a, q_b]^\mathrm{T}$, respectively. The constraints are $X = \{ x \in {R^4}:[0.2,0.2,0.2,0.2] \leqslant x \leqslant [1.36,1.36,1.30,1.30]$ and $U = {\text{ }}\{ u \in {R^2} : {\text{ }}[0,0] \leqslant u \leqslant [3.26,4]\}$. Moreover, consider the additive disturbance $\delta = [\delta_1, \delta_2, \delta_3, \delta_4]^\mathrm{T}$ with $\delta_i \in [0, 5 \times 10^{-3}]$ for tank $i$=$1,2,3,4$. By the first principle, the plant energetic consumption of the tank system can be measured by the economic cost
\begin{equation}
L^e_k = \left(q_a + 0.5 q_b^2\right) + \frac{30 V_{\min}}{S (h_1 + h_2)} + \varpi
\end{equation}
where $V_{\min} = 0.012$ and $\varpi \in [0, 2]$ satisfies uniform distribution. Note that in practice, the economic cost (54) is generally unknown. Then the control goal of the tank system aims to adjust the water levels of the four tanks from the initial state $x_0 = (0.5, 0.51, 1, 0.32)$ to the neighborhood of the equilibrium point $(x_s, u_s) = (0.7834, 0.5640, 1.0681, 0.3274, 1.1437, 2.5080)$ under the disturbance while minimizing the plant energetic consumption in the presence of constraints.

In simulation, the regressor defined in (7) with eight kernels is adopted to learn the economic cost(58). We use the polynomials and Gaussian kernel functions in the mixed kernel, i.e., 
\begin{equation}
\phi^i(w) = e^{-(w - w_i)^2 / 2\sigma^2} + \sum_{j=1}^2 \left( 0.1 w^{i^\mathrm{T}} \cdot w^i + c \right)^j,
\end{equation}
where the parameters $\sigma = 1.2247, c = 0.05$, and the initial weights are generated by a normal distribution and updated online. To evaluate the learning cost, five experiments of learning iterations are conducted, where each is initialized from the same starting point and consists of 100 time steps. Fig. 7 shows these learning results, where the upper plot 7(a) shows the evolution of the initial weights over iterations. The middle plot 7(b) gives the learning errors and the lower plot 7(c) shows the iterative refinement of the cost, combined with the EMPC. From Fig. 7, it is known that the weights gradually converge to the local optimum as online data are collected, which decreases the prediction errors and improves the economic performance across iterations. Thus, the utilization of online cost learning can effectively enhance the economic performance.

Furthermore, the function (56) is utilized as the auxiliary stable cost function with $\mathbf{R} = I_2$, $\mathbf{Q} = I_4$, and the terminal penalty matrix $\mathbf{P}$ is computed using the function `dlqr'. Then the parameters of terminal sets are estimated as $\alpha_p = 0.0320, \alpha = 0.0210$ and $d = 0.0043$. We choose $N_0 = 22$ and fix the kernel weights after the fifth iteration. Moreover, we use four different parameter combinations (PC) of $(\upsilon_k, \sigma_k)$ to validate the effect of horizon changes, i.e.
{\small
\begin{equation}
\begin{aligned}
&\text{PC1:} \begin{cases} \upsilon_k = 0 \\ \sigma_k = 0 \end{cases}, &
&\text{PC2:} \begin{cases} \upsilon_k = 0.2/\sqrt{k} \\ \sigma_k = 0 \end{cases} \\
&\text{PC3:} \begin{cases} \upsilon_k = 1 \\ \sigma_k = 0 \end{cases}, &
&\text{PC4:} \begin{cases} \upsilon_k = 0.2/\sqrt{k} \\ \sigma_k = \lceil 0.01(k-10) \rceil \end{cases}
\end{aligned}
\end{equation}}

The corresponding simulation results are displayed in Fig.8, where the plot 8(a) shows the time evaluation of the prediction horizon, the plot 8(b) displays the computational time of online solving the FHEOC problem at each time step. In the plot 8(c), we let $\Pi(x_k, N_k) = {N_k}d + \lambda a$ when (12f-1) is activated for the convenience of drawing. Plots 8(d)-8(g) show the time evaluations of the contractive sequence, control input $u_1$, closed-loop state $x_2$, $x_3$ and economic stage cost, respectively, and the plot 8(h) illustrates the average economic performance achieved by the proposed EMPC. In Fig. 8, The blue solid, green dash-dotted, orange dash, and black dotted lines correspond to PC1, PC2, PC3, and PC4, respectively. 

From plot 8(a), one can see that by tuning the parameters $(\upsilon_k, \sigma_k)$, the horizon is controlled as the form of constant horizon, fastest descent, slow descent, and first increasing and subsequent decreasing. As the horizon is linearly proportional to the number of constraints and decision variables in FHEOCs, the computational time tendency is similar to the horizon, which is pictured in plot 8(b). Note that the curves of the computation time fluctuate due to different initial points of optimization and activated constraints. In plot 8(c), the contractive sequence $\Pi(x_k, N_k)$ decreases steadily if the horizon is non-decreasing but may increase if the horizon is decreasing (e.g., $k=7$). Due to the non-dissipativity of the system setting, larger $\Pi(x_k, N_k)$ allows for aggressive control inputs in plot 8(d). This results in the system state deviating from the equilibrium point, as displayed in plots 8(e) and 8(f), but improving transient performance as shown in plot 8(g). Moreover, from the plot 8(h), it is known that the PC3 exhibits the best average economic performance. This results from the fact that the horizon maintains the decreasing condition for the longest time, thus achieving the most significant improvement in transient performance. In fact, among the four parameter combinations, the longer the decreasing horizon is maintained, the better the average economic performance is. However, this comes at the cost of the time delay of the closed-loop system to ensure the ISS property, which explains why the trajectories with constant horizon in plots 8(d) and 8(f) converge the fastest to the neighborhood of the steady state while PC3 is the slowest to converge but has best performance amongst all PCs.

\section{CONCLUSION}
This paper proposed a novel varying horizon leaning robust EMPC without terminal constraints for constrained nonlinear systems with bounded disturbances and unknown economic costs, where the horizon was adjusted online to balance the computational burden, economic performance, and convergence speed. Leveraging the multi-objective approach decoupling closed-loop stability and economic performance, a horizon-dependent contraction constraint was designed to ensure the system converges to the neighborhood of any user-specified steady state. Then an iterative procedure was presented to online adjust the horizon, strictly ensuring constraint satisfaction and recursive feasibility. The feasibility and ISS properties were proved under some sufficient assumptions. The advantages of the proposed VHEMPC were verified by the comparison results on a CSTR and a four-tank system. 

The future work will focus on considering time-varying steady-state points under learning economic costs as well as reducing the conservativeness of the terminal-free tube method.

\section{ACKNOWLEDGEMENT}                               
This work is supported by the National Natural Science Foundation of China (62173303, U24A20270).

\section{APPENDIX}         
\textbf{Proof of Lemma 2.} In Parts 1 and 2 below, we respectively prove that, with $N_{k+1} = N_k$, the conclusion holds under cases A.1) and A.2), and then explain the case of $N_{k+1} > N_k^*$ in Part 3.

\textbf{Part 1.} Case A.1) implies that $x_k \in X_T$, $x^*_{N_k|k} \notin X_T$, and thus
\begin{equation}
V_a^e({x_k},{N_k})\mathop  \geqslant \limits^{(11),(12a)} \lambda a + {L_a}({x_k},u_{0|k}^*).
\end{equation}
Moreover, there is   
\begin{equation}
\begin{gathered}
  {E_a}({x_{k + 1}}) = {E_a}\left( {f\left( {{x_k},\pi ({x_k})} \right) + {\delta _k}} \right) \hfill \\
  \mathop  \leqslant \limits^{(14),(15b)} {E_a}({x_k}) - {L_a}({x_k},\pi ({x_k})) + {c_E}\bar \delta \\
   \mathop  \leqslant a + {c_E}\bar \delta  \mathop  \leqslant \limits^{(28)} {a_p} \hfill \\ 
\end{gathered} 
\end{equation}
which means ${x_{k + 1}} \in {X_p}$. As a result, the candidate solution (22) ensures that the entire state predictive trajectory (24) is within the set $X_p$. Therefore, the constraints (12b)-(12e) trivially hold. 

Further, by Assumption 3, we have 
\begin{equation}
\begin{gathered}
  {J_a}({x_{k + 1}},\mathbf{\bar u}({N_{k + 1}}),{N_{k + 1}}) \hfill \\
  \mathop  = \limits^{{\text{(10)}}} \sum\nolimits_{i = 0}^{{N_{k + 1}} - 1} {{L_a}\left( {{x_{i|k + 1}},\pi ({x_{i|k + {\text{1}}}})} \right) + } \lambda {E_a}({x_{{N_{k + 1}}|k + 1}}) \hfill \\
  \mathop  \leqslant \limits^{(14)} \lambda {E_a}({x_{k + 1}}) \leqslant \lambda a \leqslant V_a^e({x_k},{N_k}) - {L_a}({x_k},u_{0|k}^*) \hfill \\
   \leqslant V_a^e({x_k},{N_k}) - \mu {L_a}({x_k},u_{0|k}^*) + \bar \delta {\xi _{{N_k}}}. \hfill \\ 
\end{gathered}
\end{equation}

\textbf{Part 2.} The satisfaction of (12b)–(12d) is trivial. From Lipschitz condition (3) and tightened constraint sets in (13), we have
\begin{subequations} 
\begin{equation}
\left\| \overline{x}_{i-1|k+1} - x^*_{i|k} \right\| \leq L_f^{i-1} \bar{\delta}, \quad \forall i \in \mathbf{I}_1^{N_k^*}
\end{equation}
\begin{equation}
\overline{x}_{i-1|k+1} \in L_f^{i-1} W \oplus X_i \subseteq X_{i-1}, \quad \forall i \in \mathbf{I}_1^{N_k^*}.
\end{equation}
\end{subequations}
Moreover, we have
\begin{equation}
\begin{aligned}
& E_a \left( \overline{x}_{t^*_{N_k+1} - 1|k+1} \right) \leq E_a \left( x^*_{t^*_{N_k+1}|k} \right) + c_E L_f^{N_k^* -1} \bar{\delta} \\
& \leq a + c_E L_f^{N_k^* -1} \bar{\delta} \mathop  \leqslant \limits^{(28)} a_p
\end{aligned}
\end{equation}
which implies that $\overline{x}_{t^*_{N_k+1} -1|k+1} \in X_p$ and $\overline{x}_{i-1|k+1} \in X_T$, $\forall i \in \mathbf{I}_{t^*_{N_k+1} + 1}^{N_{k+1}}$ by (16).

Thus, the (12e) is satisfied. Furthermore, with the construction in (22) under case A.2), there is
\begin{equation}
\begin{aligned}
& J_a(x_{k+1}, \bar{\mathbf{u}}(N_{k+1}), N_{k+1})  \\
&\leq \sum_{i=0}^{t^*_{N_{k+1}} - 1} L_a(\overline{x}_{i|k+1}, u^*_{i|k})  \\
& + \sum_{i=t^*_{N_{k+1}}}^{N_{k+1} - 1} L_a(\overline{x}_{i|k+1}, \pi(\overline{x}_{i|k+1})) 
+ \lambda E_a(\overline{x}_{N_{k+1}|k+1}) \\
&\leq \lambda E_a(\overline{x}_{t^*_{N_{k+1}}|k+1}) 
+ \sum_{i=0}^{t^*_{N_{k+1}} - 1} L_a(\overline{x}_{i|k+1}, u^*_{i|k}).
\end{aligned}
\end{equation}
If $t^*_{N_{k+1}} = N_k^*$, that is $x^*_{N_k|k} \in X_T$, it follows that
{\small
\begin{equation}
\begin{aligned}
& \lambda E_a(\overline{x}_{N_k^*|k+1}) 
+ \sum_{i=0}^{N_k^* - 1} L_a(x_{i|k+1}, u^*_{i|k+1})  \\
&\leq \lambda E_a(\overline{x}_{N_k^*|k+1}) - \lambda E_a(x_{N_k^*|k}) 
+ L_a(x_{N_k^*|k}, u^*_{N_k^*|k+1}) \\
&\quad + \sum_{i=0}^{N_k - 1} \left[ L_a(\overline{x}_{i|k+1}, u^*_{i|k+1}) - L_a(x_{i|k}, u^*_{i|k}) \right] \\
&\quad + V_a^e(x_k, N_k) - L_a(x_k, u^*_{0|k}) \\
&\leq V_a^e(x_k, N_k) + \bar{\delta} \xi_{N_k} - \kappa L_a(x_k, u^*_{0|k}).
\end{aligned}
\end{equation}}

Otherwise, $t^*_{N_{k+1}} < N_k^*$ implies $x^*_{N_k^*|k} \notin X_T$, it follows that
{\small
\begin{equation}
\begin{aligned}
& \lambda E_a(\overline{x}_{t^*_{N_{k+1}}|k+1})  
+ \sum_{i=0}^{t^*_{N_{k+1}} - 1} L_a(\overline{x}_{i|k+1}, u^*_{i|k+1})  \\
&\leq \lambda a + \sum_{i=0}^{t^*_{N_{k+1}} - 1} L_a(x^*_{i|k+1}, u^*_{i|k+1}) \\
&\leq \lambda a + \sum\limits_{i = 0}^{t_{{N_{k + 1}}}^* - 1} {{L_a}(x^*_{i|k+1},u_{i + 1|k}^*)} + \sum_{i=0}^{t^*_{N_{k+1}} - 1} \bar{\delta} c_L \cdot L_f^i \\
&\leq \lambda E_a(x^*_{N_k^*|k}) 
+ \sum_{i=0}^{t^*_{N_{k+1}} - 1} L_a(x^*_{i|k+1},u^*_{i+1|k}) + \bar{\delta} \xi_{N_k} \\
&\leq V_a^e(x_k, N_k) - L_a(x_k, u^*_{0|k}) + \bar{\delta} \xi_{N_k}.
\end{aligned}
\end{equation}}
\textbf{Part 3.} Note that the tail elements of candidate sequence (22) are formulated by $\pi$  via the state inside $X_T$. Thus, (31) can be obtained by iterating (14). $\Box$

\normalsize
\bibliographystyle{apalike}     
\bibliography{references}    


\end{document}